# Magnetic field induced resistance properties at filling factor 5/2 consistent with non-Abelian e/4 quasiparticles in multiple sized interferometers

R.L. Willett, Bell Laboratories, Alcatel-Lucent
L.N. Pfeiffer, K.W. West, Princeton University

ABSTRACT:

Non-Abelian e/4 quasiparticles at 5/2 filling factor in a correlated two-dimensional electron gas have a proposed specific property in an interference measurement of their edge propagation: encircling an even number of localized e/4 quasiparticles allows expression of e/4 Aharonov-Bohm (A-B) oscillations, but suppression of these oscillations will occur with an odd number encircled. This picture is tested explicitly here in multiple interferometers of different areas. The encircled localized e/4 quasiparticle number near 5/2 filling factor is changed by sweeping B-field, and oscillations are observed in resistance near 5/2 of period specific to that device area for each interferometer. The product of the measured interferometric area of each device and the respective 5/2 resistance oscillation period is found to agree with the expected flux quanta addition needed for parity change in the localized e/4 number. This result shows a highly specific non-Abelian property of the quasiparticle excitations at 5/2 filling expressed in multiple interferometers.

The origin of the fractional quantum Hall effect [1] at 5/2 filling factor [2] has been a long standing open issue. This quantum Hall state anomalously forms at that even denominator filling factor and so has demanded an explanation beyond both the Laughlin state wavefunction and its hierarchy of fractional states [3]. This problem is of particular interest in physics in general with its proposed potential to be due to a spin polarized p-wave pairing of composite fermions [4], the Moore-Read state [5]. This potential state would have e/4 charge excitations that display non-Abelian statistics [5-7] due to their Majorana zero mode operators [6, 8]. The challenge experimentally has been to determine if the e/4 charges demonstrate non-Abelian properties: a particular picture has been predicted to test this possibility using interferometry that allows observation of the Aharonov-Bohm (A-B) effect of the charged quasiparticles and controlled encircling of quasiparticles localized in the interferometer [9-11]. In this prescription e/4 particle exchange with an even or odd number dictates whether e/4 oscillations are expressed in the Aharonov-Bohm (A-B) measurement; if an even number of e/4 quasiparticles are encircled the e/4 A-B oscillations are expressed, but are suppressed if an odd number is encircled.

Past results [12-13] have established interferometric observation of e/4 quasiparticles at 5/2 filling factor and properties of alternation and interchange: in a double point-contact interferometer A-B oscillations corresponding to an e/4 charge are observed alternately with e/2 period oscillations [14-16], consistent with the interferometer alternately enclosing an even/odd number of localized e/4 quasiparticles. This alternation can be affected with addition of B-field to add one localized quasiparticle in the encircled area, where-in the e/4 and e/2 periods are observed to interchange [13]. These experiments were focused on device perimeter delineating gate sweeps to change the enclosed interferometer area and the consequent determination of e/4 and e/2 charge properties.

Here we examine e/4 A-B oscillations at 5/2 in a series of interferometers, with substantial variation in their active areas. We test the most specific non-Abelian property, turning on and off this expression of the e/4 oscillations by a fundamental process; changing B-field in quantum Hall effect systems changes the localized quasiparticle number. In the measurements presented here at 5/2 filling factor, e/4 oscillations in alternation with e/2 periods are observed in the various devices confirming the presence of the fundamental 5/2 excitations. The active interferometric area of each is measured through A-B oscillations at integer filling factors for fixed gate voltages and simple B-field sweep. At the same gate voltages this sweep of B-field near 5/2 reveals small period resistance oscillations with periods specific to each device size. These periods should correspond to the changing parity of the enclosed quasiparticle number, as the e/4 oscillations are expressed and then suppressed as B-field is changed. If this is actually the mechanism of the B-field oscillations at 5/2, the product of this observed period and the device area should equal the change in the value of the flux quantum number necessary to add parity changing localized e/4. This measured product is indeed found to be consistent with the expected flux quanta addition for all of the devices tested, covering a range of areas. This is a highly specific test of the non-Abelian nature of the e/4 excitations.

The top gate structures that define the multiple interferometers fabricated on the high mobility heterostructure wafer used in this study are shown in Fig. 1. Two interfering edge currents result from backscattering at gate sets 1 and 3. The longitudinal resistance $R_L$ is measured with labeled contacts a through d in the electron-micrograph by the voltage drop from contact a to d, with current driven from b to c. Prior to charging the top gates the samples can be briefly illuminated to enhance mobility and to provide different sample preparations since the illumination changes the charge localizing profiles [12-13, 17]. The two standard top gate designs shown in electron micrographs in the Figure are labeled with device dimension parameters x and y adjusted to produce three separate samples with ratios of areas roughly 3:2:1. The range of functional areas in these devices is also defined by the continuous range of gate voltages applied, ultimately resulting in a range of areas from ~ 0.1 to 0.6$\mu$m$^2$. The active area of the interferometer is measured via A-B resistance oscillations in B-field sweep measurements near integer filling over the entire filling factor range, with oscillation period in B related to the device area via $\Delta B_1 = \phi/A = 40$Gauss-$\mu$m$^2$/A. Further description of methods, data analysis, specifics of the large lateral 2D gas depletion, and issues of potential Coulomb dominated properties are provided in the supplemental information.

The results below will show first e/4 and e/2 A-B oscillations near 5/2 filling factor as a function of side gate voltage sweep for each of three devices, with the period scaling there determined by similar measurements at integer and 7/3 filling factors; these results are consistent with past findings of similar gate sweep studies [12-13], and establish the presence of the known e/4 quasiparticle properties in the varied area devices of this study.

The principal focus of this study is then presented in its two essential parts; B-field sweeps at 5/2 showing B-field induced small-period oscillations in resistance, and at equivalent top gate values the corresponding area determinations for each preparation of three different interferometers. The product of the area and those 5/2 resistance oscillation periods for each preparation is then shown to be consistent with the added flux quanta needed to induce the quasiparticle number parity change.

Standard transport through each of the three devices is shown in Fig. 1, demonstrating prominence of the 5/2 minimum in longitudinal resistance, and preservation of the complex of fractions (8/3 and 7/3) around 5/2 typical for a 2D electron system of high mobility.

A-B oscillations in resistance versus side gate voltage change corresponding to e/4 charge are present in each of the devices tested here, with these results shown in Fig. 1 (right hand column). The results of this figure demonstrate the e/4 properties established in prior studies [12-13] are present throughout the extended range of device areas used in this study. The periodicity marked by vertical lines in each right hand column figure corresponds to that for e/4 charge near 5/2 derived from similar measurements at integer and 7/3 filling factors, from the relationship P~h/(e*B), where e* is the quasiparticle charge and B is the field at which the oscillation is measured. While the phase is adjusted, a good correspondence to e/4 is achieved from those measurements at other filling factors and B-fields. Also apparent in the results of Fig. 1 is the alternation between periodicities of e/4 and e/2, as observed in previous studies [12-13]. This alternation corresponds to the side gate excursion enclosing an even number of localized e/4 quasiparticles for expression of e/4 periods, and this expression is suppressed when further side gate excursion encloses an odd number of localized quasiparticles. When an odd number of quasiparticles is enclosed, the background A-B oscillations from pervasive Abelian e/2 quasiparticles are observed [14-16]. The precise pattern of this alternation is dependent upon the spatial pattern of the native localized quasiparticle population, and as such is expected to be different for different excursions of the same side gate and for different samples preparations.

This alternation is one of the signatures of the non-Abelian nature of the e/4 excitation, and it can be tuned in a specific way with change in B-field. By addition of magnetic field sufficient to add just one localized quasiparticle, the alternation pattern can be changed such that where e/4 oscillations occurred, now e/2 will be present, and where e/2 were present, e/4 will now be expressed. This is the basis of the principal finding of this study: by adding a single localized quasiparticle through magnetic field change, the parity of the enclosed number is changed, and the e/4 expression or suppression is reversed, causing an interchange of those two periods, with a resultant change in the resistivity. Therefore a continuous change in B-field near 5/2 should cause resistance oscillations with a period that represents two parity changes, reflecting the process of addition of one then two localized e/4 quasiparticles. That period is explicitly dependent upon the area of the device.

This method of changing only the B-field but maintaining the top gate voltages is a particularly important factor in measurements on these samples. In previous studies, as is true here, the gate voltages are allowed to stabilize over long time periods (hours) in order to avoid charge drift, and so in swept gate measurements long sweep times are needed: in the $\Delta$Vs data of Fig. 1, each trace of 100mV excursion required 6 hrs. In a simple B-sweep measurement, the gate voltages can be applied and allowed to equilibrate prior to the sweep, with subsequent B-field dependent resistance not affected by gate charging factors.

In this picture of altering the localized quasiparticle number through B-field change, if a single gate voltage is chosen near 5/2 at a peak or minimum in the e/4 oscillation, by sweeping B-field this resistance value will be alternately expressed or suppressed, resulting in an oscillation in resistance. The period of this oscillation is specifically defined in that it would correspond to adding two localized quasiparticles to that enclosed area through the following counting argument: At 5/2, there are five electrons per two flux quanta, and four e/4 quasiparticles per electron, or $(5e/2\phi)(4e^*/e)=10e^*/\phi$. Therefore to add two localized e/4 quasiparticles to an area A, the necessary added field (which we will call $\Delta B_2$) is $2(0.1\phi)/A= 2(0.1)(40\text{Gauss-}\mu m^2)/A=8\text{Gauss-}\mu m^2/A$, or $\Delta B_2$ x Area = $8\text{Gauss-}\mu m^2$. (Note that the active interferometric device area is simply measured through the A-B oscillations present near integer filling factors, with area $A=\phi/\Delta B_1$.)

Demonstration of just such prescribed small period resistive oscillations corresponding to multiple parity changes in the enclosed e/4 quasiparticle number is shown in Fig. 2, top panel, for one of the sample preparations of device area 2. In this model of parity change by B-field sweep, the five oscillation periods shown represent 10 parity changes. The B-field range near 5/2 filling factor where this data is taken is marked in the overall $R_L$ trace. The distinct resistive oscillations here are from an average of ten B-field sweeps (black trace; the red trace is a coarse smoothing of the data), with the 5/2 resistive oscillation period $\Delta B_2$ in this preparation roughly 22 Gauss. The set of oscillations are measured with no adjustments in the voltages of the quantum point contacts or central top gates.

Also displayed in Fig. 2 are two sets of A-B oscillations taken near integer filling 2 and 3. The active area of this preparation is measured by this periodicity, $\Delta B_1$. The period $\Delta B_1$ is found to be similar near multiple different integer filling factors for each device, consistent with this being an A-B oscillation. As A-B oscillations, the period is related to the active interferometer area, $\Delta B_1$ = 40 Gaus-$\mu m^2$/A: the data of Fig. 2 with period $\Delta B_1$ roughly 110 Gauss result in an area of about $0.36\mu m^2$.

Measurement of these oscillation sets (period $\Delta B_2$ and the corresponding $\Delta B_1$) was repeated for multiple interferometric areas in order to examine whether this oscillation is consistent with the picture of addition of localized non-Abelian e/4 quasiparticles inducing parity change in the interference since the $\Delta B_2$ period should change with area in this model. From the three devices used and the multiple preparations and gate values employed, the measured $\Delta B_1$ periods show active areas ranging from <0.1 $\mu m^2$ to ~ 0.6 $\mu m^2$. Four additional pairs of the 5/2 parity oscillations with period $\Delta B_2$ and their corresponding area A-B oscillations with period $\Delta B_1$ are shown in Fig. 3, extracted from different devices and/or sample preparations. All the $\Delta B_2$ traces are measured with the low field end of the B-sweep at or near 5/2, and the data are all taken on the high B-field side of 5/2. Note that as device area increases (decrease in $\Delta B_1$) the parity oscillation period ($\Delta B_2$) decreases. A total of 11 sample preparations were tested using the three different devices.

Fig. 4 shows the principal summary result of this study. The inset shows the 5/2 parity oscillations $\Delta B_2$ versus area derived from $\Delta B_1$ for these different preparations. Good agreement with $\Delta B_2 = 8\text{Gauss-}\mu m^2/A$ is shown. The overlay plot shows the product of area and B-field period $\Delta B_2$ versus area, with the level of 8 Gauss-$\mu m^2$ marked. The data are distributed around this number, but are in coarse agreement with this expected value. The data show that two periods of ostensibly different origin produce a single prescribed value of flux. This indicates that in multiple devices with different gate values, the product of interferometer area and resistance oscillation period $\Delta B_2$ at 5/2 is consistent with the expected flux change necessary to add two localized e/4 quasiparticles within this model of non-Abelian e/4.

In examining the parity oscillations with period $\Delta B_2$, it might be expected that the oscillatory property should be closer to a square wave than to a sine wave; the resolution of the measurement does not offer that distinction. In each experimental sample configuration the gate voltages were allowed to equilibrate over a large time period (> 20 hours) before data was collected, and only B field change is used.

This experiment is the necessary complement to prior measurements [12-13] examining A-B derived charges at 5/2 using gate voltage change to alter the active interferometer area A. By sweeping B only in these multiple area devices, the experimental limitation in the prior work of slow gate charging is avoided. More importantly, the most fundamental process within this new statistics of e/4 is addressed. The theoretically predicted $R_L$ oscillation upon sweeping B near 5/2 is exposed, and the oscillation period is found to quantitatively agree over a range of device areas with the value of flux quanta necessary to add two localized e/4.

In this study a specific property of non-Abelian e/4 quasiparticles is extracted for a series of interferometric devices of different sizes. It is shown that resistance oscillations are present near 5/2 filling factor, and they are consistent in period with the prescribed B-field necessary to add localized quasiparticles to their respective different areas, inducing expression/suppression of A-B oscillations, in a simple model of this e/4 non-Abelian excitation. This experiment demonstrates that a single magnetic field sweep and resistance measurement through these interferometers is able to reveal crucial aspects of the consistency of e/4 properties and predicted non-Abelian characteristics.

**References:**

(1) D.C. Tsui, H.L. Stormer, A.C. Gossard, Two-dimensional magnetotransport in the extreme quantum limit, Phys. Rev. Lett. **48**, 1559 (1982).

(2) R.L. Willett et al, Observation of an even-denominator quantum number in the fractional quantum Hall effect, Phys. Rev. Lett. **59**, 1776 (1987).

(3) R.B. Laughlin, Anomalous quantum Hall effect: an incompressible quantum fluid with fractionally charged excitations, Phys. Rev. Lett. **50**, 1395 (1983).

(4) see article by Jainendra Jain, Physics Today, April 2000 and references there in.

(5) G. Moore, N. Read, Nonabelions in the fractional quantum Hall effect, Nucl. Phys. B**360**, 362 (1991).

(6) C. Nayak, F. Wilczek, A Chern-Simons effective field theory for the Pfaffian quantum Hall state, Nucl. Phys. B**479**, 529 (1996).

(7) P. Bonderson, V. Gurarie, C. Nayak, Plasma anology and non-Abelian statistics for Ising-type quantum Hall states, Phys. Rev. **B**83: 075303 (2011).

(8) N. Read, D. Green, Paired states of fermions in two dimensions with breaking of parity and time-reversal symmetries and the fractional quantum Hall effect, Phys. Rev. B**61**, 10267 (2000).

(9) A. Stern, B.I. Halperin, Proposed Experiments to Probe the Non-Abelian ν=5/2 Quantum Hall State, Phys. Rev. Lett. **96**, 016802 (2006).

(10) P. Bonderson, A. Kitaev, K. Shtengel, Detecting non-Abelian statistics in the ν=5/2 fractional quantum Hall state, Phys. Rev. Lett. **96**, 016803 (2006).

(11) S. Das Sarma, M. Freedman, C. Nayak, Topologically Protected Qubits from a Possible Non-Abelian Fractional Quantum Hall State, Phys. Rev. Lett. **94**, 166802-166805 (2005).

(12) R.L. Willett, L.N. Pfeiffer, K.W. West, Measurement of filling factor 5/2 quasiparticle interference with observation of e/4 and e/2 period oscillations, Proc. Natl. Acad. Sci. **106**, 8853 (2009).

(13) R.L. Willett, L.N. Pfeiffer, K.W. West, Alternation and interchange of e/4 and e/2 period interference oscillations consistent with filling factor 5/2 non-Abelian quasiparticles, Phys. Rev. B **82**, 205301, (2010).

(14) X.Wan, Z. Hu, E. Rezayi, K. Yang, Fractional quantum Hall effect at ν=5/2: Ground states, non-Abelian quasiholes, and edge modes in a microscopic model, Phys. Rev. **B** 77,165316 (2008);

(15)W. Bishara, P. Bonderson, C. Nayak, K. Shtengel, J. Slingerland, The non-Abelian interferometer, Phys. Rev. B**80**, 155303 (2009)

(16) W. Bishara, C. Nayak, Odd-even crossover in a non-Abelian ν=5/2 interferometer, Phys. Rev. B**80**, 15304 (2009).

(17) R.L. Willett, M.J. Manfra, L.N. Pfeiffer, K.W. West, Confinement of fractional quantum Hall states in narrow conducting channels, Appl. Phys. Lett. **91**, 052105-052107 (2007).

**Figure Captions:**

Figure 1: Interferometer design, magneto-transport through the devices, and gate sweep induced A-B oscillations at 5/2 filling factor. The two electron micrographs demonstrate the fundamental design layouts for the interferometers of this study; each is comprised of pairs of quantum point contacts labeled 1 and 3, and a third gate 2 that can independently adjust area A. Transport contacts are labeled a-d. From these base designs three devices were produced by variation in dimensions labeled x and y, with dimension pair products in each of $x_1y_1$, $x_2y_2$, $x_3y_3$ resulting in ratios of roughly 1:2:3 in the lithographic areas. The interferometric functional area A is determined by the base device lithographic design and adjustment of the gate voltages, resulting in areas ranging from ~ 0.1$\mu m^2$ to 0.6$\mu m^2$ as derived from A-B oscillation periods $\Delta B_1=\phi/A=40$Gauss-$\mu m^2$/A. All devices used here are from the same heterostructure wafer, with representative bulk transport shown in the lower left panel. The central figure column shows representative transport through each device; note prominence of the 5/2 minimum in this longitudinal resistance for each. The right hand column shows longitudinal resistance change with side gate (2) sweep at filling factor 5/2; each device demonstrates oscillations consistent with A-B effect at periods for charge e/4. The marked vertical lines of these periods are derived from similar measurements at integers and 7/3, defining the period corresponding to e/4

charge. In each device the previously observed [12-13] alternation of e/4 and e/2 periods is affirmed in large side gate voltage excursions. Temperature in all data is ~ 25mK.

Figure 2: The fundamental finding of this study is the presence of the small B-field oscillation near 5/2 filling (top panel) corresponding to addition of e/4 quasiparticles to the interferometer area A, which induces alternately expression or suppression of the interference consistent with non-Abelian e/4 quasiparticles. This B-field period $\Delta B_2$ is dependent upon the area A, and both are measured in simple B sweep over the filling factor range 2 to 3, with coarse $R_L$ displayed in the middle panel. As shown in the figure, in proximity to 5/2 a small oscillation period of ~ 22G is observed (corresponding to addition of two localized e/4 quasiparticles for the full period), and the active device area is derived from A-B oscillations of period $\Delta B_1$ present near integer filling factors (lower panels). For each preparation in the study multiple integer filling A-B oscillation sets can be used to determine each area. The data here are from a single preparation in device of area 2, with temperature ~25mK.

Figure 3: pairs of area determining integer filling factor A-B oscillations (right hand column) and their corresponding transport near 5/2 showing small period oscillations consistent with expression/suppression of non-Abelian e/4 interference. In the model of these oscillations as manifestations of adding localized quasiparticles, the filling factor 5/2 and the device area A determine the B-field necessary to add one quasiparticle: at 5/2, $5e/2\phi \times 4e^*/e = 10e^*/\phi$ , or $0.1\ \phi/e^*$; to add one quasiparticle $e^*$ , B-field is changed $\Delta B_2$ (= $\Delta B(5/2)$) $\times A = 0.1\ \phi$ , where $A = \phi/\Delta B_1$, or $\Delta B_2/\Delta B_1 = 0.1$; to add two $e^*$ (for a full period) $\Delta B_2/\Delta B_1 = 0.2$, or $\Delta B_2 \times A = 8G\text{-}\mu m^2$. The black line is an average of typically eight B-field sweeps, and the red line is an adjacent average smooth of that data.

Figure 4: (inset) 5/2 periods ($\Delta B_2$) as a function of A-B measurement derived device area A and (overlay) the product of area A and 5/2 period, which for consistency with non-Abelian e/4 should be 8Gauss-$\mu m^2$. Inset: The 5/2 period drops with increasing device area over the full range of accessible areas; the solid line drawn is $\Delta B_2 = 8G\text{-}\mu m^2/A$. Overlay: The product of area and $\Delta B_2$ corresponds well to 8G-$\mu m^2$, supporting the picture of non-Abelian e/4. The blue lines on the symbols are approximations of the measurement error.

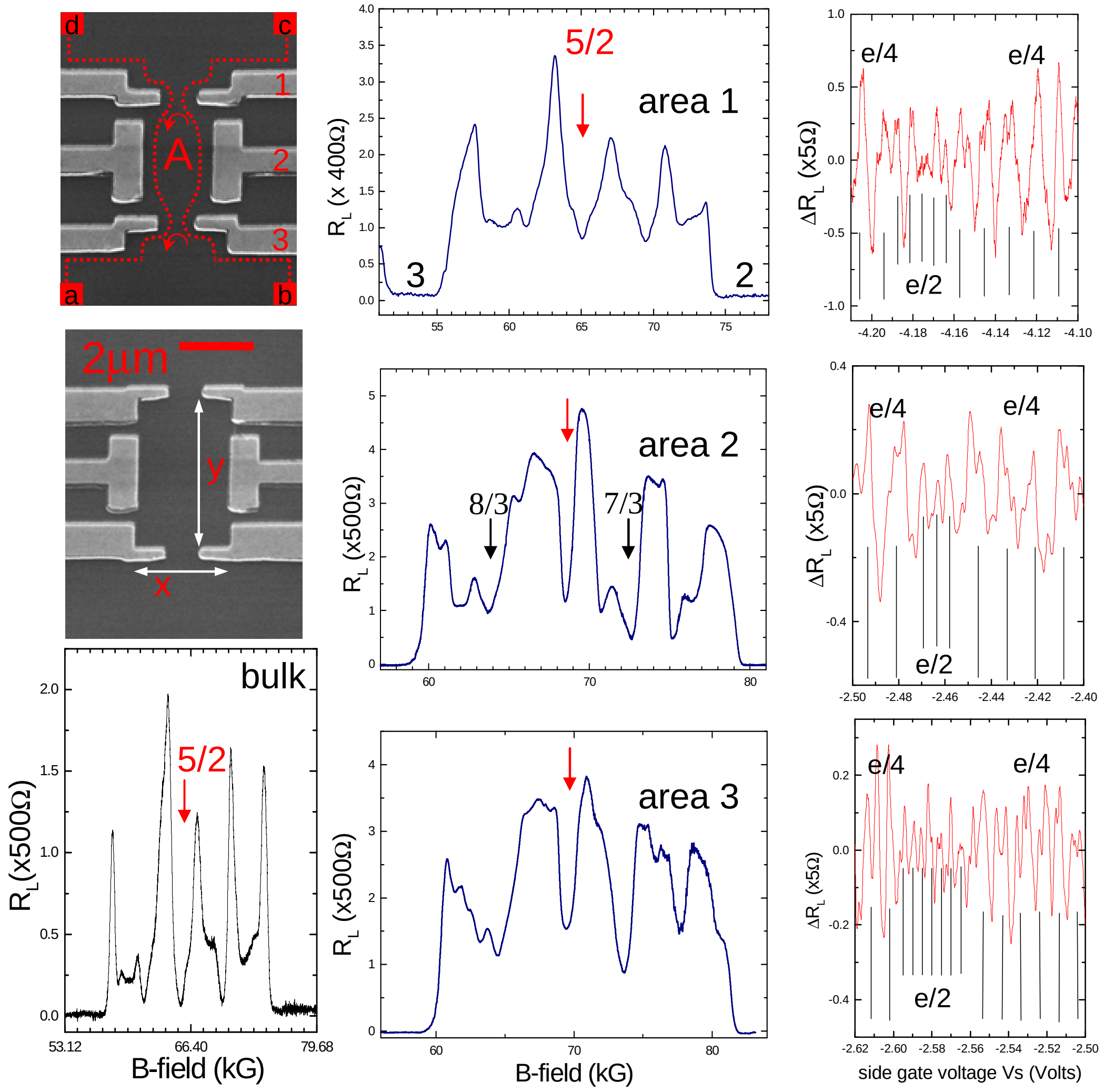

Figure 1

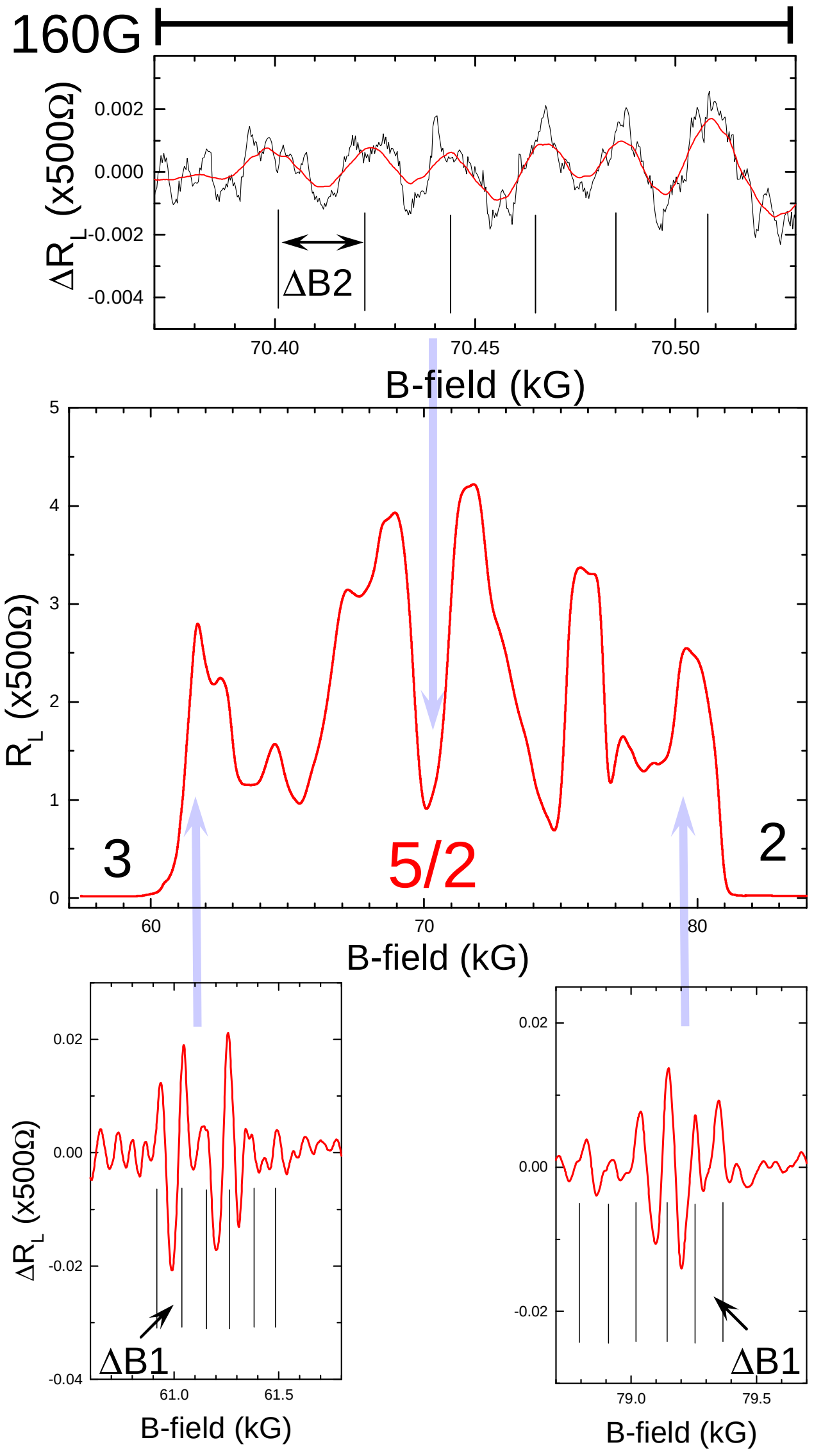

Figure 2

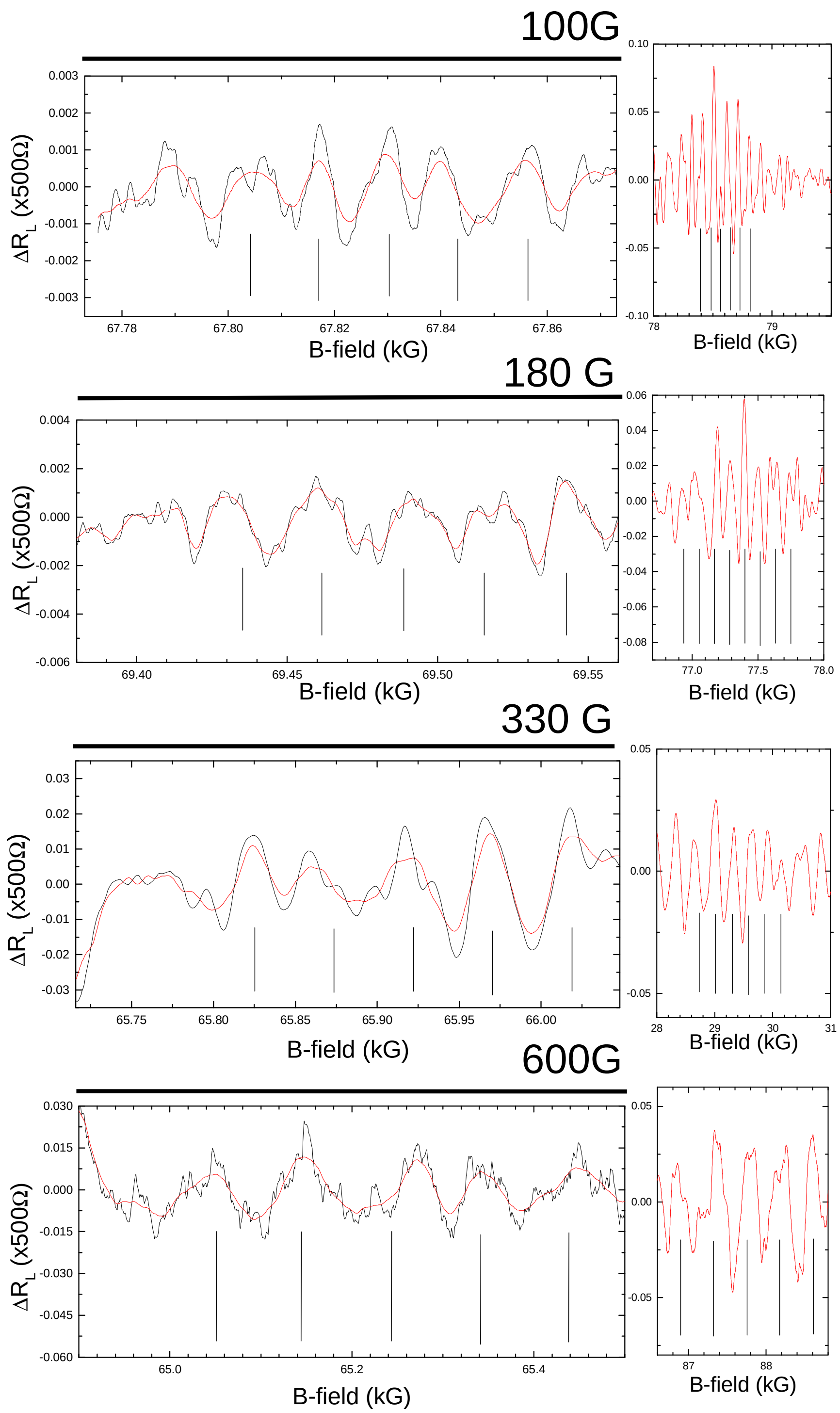

Figure 3

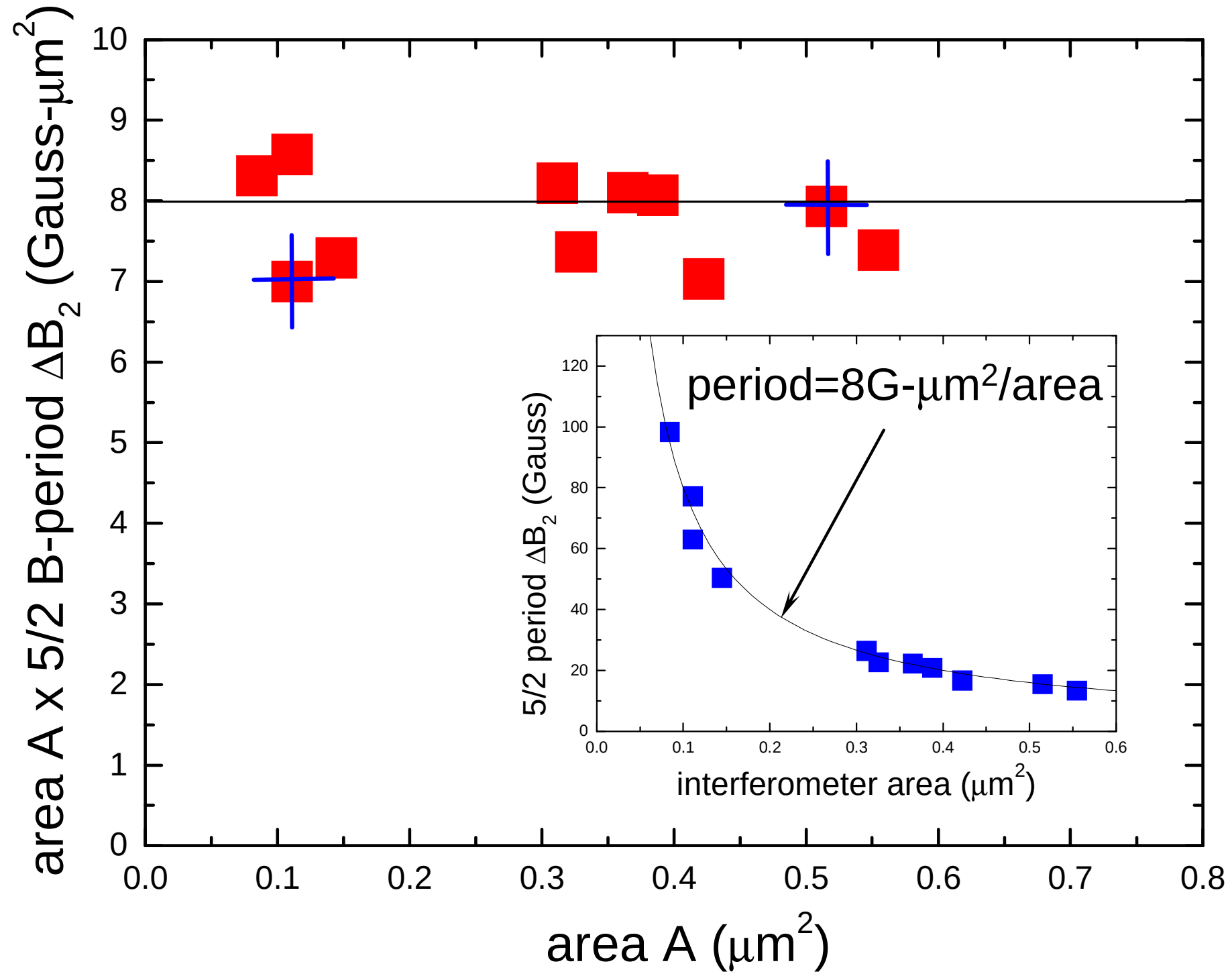

Figure 4

Supplemental Information to:

# Magnetic field induced resistance properties at filling factor 5/2 consistent with non-Abelian e/4 quasiparticles in multiple sized interferometers

R.L. Willett, Bell Laboratories, Alcatel-Lucent,
L.N. Pfeiffer, K.W. West, Princeton University

Sections:

## I) METHODS and MATERIALS:

The heterostructure wafer used in these experiments has a relatively high density of 4 x $10^{11}$ $cm^{-2}$ and high mobility of 28 x $10^6$ $cm^2$/V-sec, with the 2D electron channel 200nm below the sample surface. The top gate structures that define the multiple interferometers used here are shown in Figure 1, main text; as in samples of previous studies [1-2], a 40nm amorphous SiN layer is applied to the sample surfaces and the 100nm thick Al top gates are defined on that surface. The top gate structures are charged to operation values and typically held there for more than 20 hours before B-field sweep measurements are made. Standard lock-in techniques at low frequencies are employed in dilution refrigerators reaching base temperatures of around 20mK. From the labeled contacts a-d in the electron-micrograph of main text Figure 1, longitudinal resistance $R_L$ is measured by the voltage drop from contacts a to d, with current driven from b to c.

The samples can be briefly illuminated when the gates are not charged to enhance mobility, and so different sample preparations can be achieved as described in prior studies [1-2]. These constitute different sample preparations as the charge localizing potentials change with illumination; zero field resistance from diffuse boundary scattering is observed to be altered significantly with illumination [3].

The two standard top gate designs shown in the electron micrographs in Figure 1 are the basis for a series of interferometers. The interfering edge paths are delineated in the top photo with back scattering at the nominal quantum point contact gates marked 1 and 3. The adjusted device dimension parameters x and y are shown in the second photo. Lithographically defined device areas in three separate samples are produced using x, y pairs of the following dimensions in μms: device 1; 1.6, 2.0, device 2; 2.1, 3.0, and device 3; 2.8, 3.5, resulting in ratios of areas roughly 3:2:1. The range of functional areas in these devices is also defined by the continuous range of gate voltages applied, and the combined variation of lithographic design and voltages resulted in a range of active device areas from <0.1 to ~0.6$\mu m^2$.

In practice, the functional area of the device is substantially smaller than the lithographically defined area since the depletion mechanism is due solely to electrostatic depletion with a large lateral component. An important limitation to this interference measurement is that the top gates must be allowed to equilibrate over long time periods, typically many hours. This limits the effectiveness of swept gate measurements, as intrinsic charge relaxation to changing voltages must be considered. In this particular experiment, focusing on only swept B-field allows full relaxation of the devices before measurements are made.

As in our previous studies of Aharonov-Bohm oscillations near 5/2 filling factor, the value of the side gate voltage change corresponding to e/4 charge is determined by similar measurements at integral filling factors and 7/3 filling factor using the relationship $\Delta Vs \sim \phi/B = h/e^*B$. With definition of the presence of e/4 interference, the active area of the interferometer is measured by examining B-field sweep measurements over the entire filling factor range near integral filling, with oscillation period in B related to the device area via $\Delta B_1 = \phi/A = h/(eA)$.

To summarize data collection details, the two measurements employed here are reviewed; $R_L$ measurement with gate voltage sweep Vs, and the predominant measurement here of $R_L$ with B sweep. In Vs sweep measurements (Figure 1) the gate sweep rate is 100mV/hour, with lock-in time constant for $R_L$ set typically at 10 seconds. For B sweep measurements, in particular the $\Delta B_2$ and $\Delta B_1$ measurements in Figure 2 and Figure 3, the B sweep rate is at or less than 10 Gauss/minute.

For all demonstrations in this study of A-B oscillations ($\Delta Vs$ and $\Delta B_1$) and the small period oscillations near 5/2 ($\Delta B_2$), the background resistance has been subtracted. In addition, for the 5/2 small period oscillations (top panel in Figure 2, left panels in Figure 3), all the black traces are the average of typically eight or more B-field sweeps in the increasing B direction, and the red lines are simple adjacent averaging smoothes to highlight the overall periodicity present.

To specifically label the data in Figure 3, the top two panels are from device size area 3, and the bottom two panels are from device size area 1. The data in Figure 2 are from device size area 2. The data in Figure S1 are from device area 1. Again, note that the top gate voltage values can and have been adjusted for different preparations to achieve a range of active interferometer areas.

## II) RESULTS AT FILLING FACTOR 7/2

The fractional quantum Hall state at 7/2 is substantially weaker than that at 5/2, but the longitudinal resistance minimum is still apparent in transport through the interferometric device of nominal area 1 size: see Figure S1A below. Given this, we examined the magneto-resistance near 7/2 and observed a small period, low amplitude oscillation similar to that observed near 5/2: see Figure S1B.

The oscillatory period here can be compared to that of the 5/2 B-field induced oscillations within this picture of the effect occurring as a consequence of changing the parity of the enclosed quasiparticle number, thus inducing alternately expression and suppression of the e/4 A-B oscillations. The counting argument for localized quasiparticle number/changing B-field is as follows: at 7/2 filling factor, there are $7e/2\phi$ x $4e^*/e$ = $28 e^*/2\phi$ =$14e^*/\phi$. Therefore to add one e*, $\Delta B=(1\phi/14e^*)/A=(40\text{Gauss-}\mu m^2/14e^*)/A$, with A the measured device area, or A x $\Delta B$ = 2.86 Gauss-$\mu m^2$. Again, the measured period is addition of *two* localized quasiparticles, so that A x $\Delta B_2$ = 5.72 Gauss-$\mu m^2$.

Magneto-resistance near 7/2 in one of the smallest area interferometric preparations is shown in Figure S1B. Small period, small amplitude oscillations are evident, qualitatively similar to those observed at 5/2 filling. This resistance was measured in a sample preparation with area corresponding to that of the lowest panel in manuscript Figure 3. In larger devices with relatively lower amplitude interference at 5/2, such oscillations at 7/2 could not be resolved.

To compare to the 5/2 data the principal part of Figure 4 has been reproduced here in Figure S1C, with the above calculated expectation of 5.72 Gauss-$\mu m^2$ marked additionally for 7/2 area x period product. The experimental product measured here is in coarse agreement with the expected value in this simple model of non-Abelian e/4 at filling factor 7/2.

## III) AHARONOV-BOHM VERSUS COULOMB BLOCKADE/DOMINATED

A critical distinction in interferometer measurements is whether oscillations observed are due to the Aharonov-Bohm (A-B) effect or due to Coulomb effects (Coulomb blockade or Coulomb dominated)[4-5] This is of particular importance when the active area of the device is small: it has been found in previous studies [6-7] that small area devices display resistance oscillation properties consistent with Coulomb domination. In larger but similar devices [6], oscillations demonstrated A-B properties.

In the measurements of this study, in all the devices and their respective area ranges acquired through different top gate voltages, the resistance oscillations have shown consistency with A-B oscillations. This empirical consistency with A-B is clearly demonstrated in the B-field induced oscillations. Over a large range of B-field, from filling factor 6 to 2, no change in the resistance oscillation period is observed. This constant B-field period over a large filling factor range is as observed for large devices in the previous studies [6] in A-B, and distinctly different from Coulomb dominated results in smaller area samples of the same study, where B-period is proportional to B. An example of this constant period at different B-field is shown in Figure 2: two B-field traces at near filling factors two and three show the same B-field period oscillations. We see this constant period over a B-field range of more than a factor of three in this and the other samples, and this constant A-B oscillation period implies a constant active area A for the interferometer for each preparation. This empirical consistency with A-B is observed in all samples tested here, and is distinctly inconsistent with C-B/C-D.

The samples used in our studies have significant differences from samples in other studies [6-7], namely density, illumination, and no surface etch, and these differences may explain why the smaller areas here present A-B oscillations. The sample densities used here are roughly twice those in the other studies; this higher density promotes both correlation effects and screening capacity, but also a resistance to area change with B-field change. Further promotion of correlation effects occurs with the sample illumination in our studies, which is not provided in the other studies; when not illuminated, our samples show poor transport through the device with little evidence for fractional states in the second Landau level. By not etching the sample surfaces, our devices retain their high quality: surface etching as performed on other devices [7] clearly destroys their ability to support correlation effects and also reduces in an uncontrolled fashion the electron density in proximity to the gate structures.

## IV) DATA ANALYSIS

In the $\Delta R_L$ traces shown in the main text Figures the local background resistance has been subtracted to display the oscillations which are typically less than a few percent of the overall resistance. This background is an adjacent average at each data point taken using a large averaging number which allows approximation of a smooth background: the specific number of data points over which the average is taken depends on the density of original data points, which depends on the rate of either the B-field sweep or the side gate change. This smoothed background is then directly subtracted from the raw data.

## V) TRANSPORT THROUGH DEVICE; GEOMETRICAL EFFECTS AND SAMPLE ILLUMINATION

A crucial part of these studies is achieving longitudinal resistance through devices that demonstrate a confined 2D system able to support the small gapped states at 5/2, 7/3, and 8/3. Beyond demonstrating these fractions in this range of device areas, an important perspective point is that the same resistive properties in $R_L$ and $\Delta R_L$, oscillations from sweeping side gate and from sweeping B-field, are displayed over this range of device dimensions. This indicates that the fundamental properties appear to be consistent over areas differing by a factor of six.

As outlined in previous studies [1-2], a layering of amorphous SiN and then deposition of the top gates with no etching is used to preserve the high quality of the device area. To achieve the highest mobilities, these samples require brief illumination before transport measurements, and this is illumination is likewise applied to the devices before gate charging. This illumination produces high resolution of the fractional states not only in the adjacent bulk but also in the devices as demonstrated by the results for all device sizes as shown in Figures 1 and 2.

The geometrical arrangement of our contacts and the gross patterning of the top gates further emphasize the importance of these high quality transport traces for these different sample sizes as indications of the preserved correlations within the device area. As shown in the electron micrographs of Figure 1, the number of squares through which transport occurs through the devices ranges approximately from two to three. External to the devices the transport contacts to the 2D gas are arranged such that roughly one square of transport area is used. This means that the interferometric device areas are contributing a large proportion of the measured resistance. This geometry emphasizes the contribution from the device area, and high quality of the transport traces through these devices attests to the quality of the confined 2D electron system.

An important principal in these measurements is that a high quality transport trace through the device ($R_L$) is weighted by the geometrical factor of the device with respect to the bulk. Since $R_L$ is necessarily measured with both bulk and device contributions, a geometry heavily weighted to the bulk (more squares of area there than in the device) can mask a poor device 2D gas quality, showing high quality transport even though the device quality is low. This is why the geometrical weighting to the device is significant. However, if the bulk transport is good yet the transport through the device is bad, no 5/2 or 7/3 discernible in the $R_L$ measurement, then the geometry is irrelevant and the device is manifestly not supporting correlation effects.

## VI) LATERAL DEPLETION AND SMALL ACTIVE AREA VERSUS LITHOGRAPHIC AREA

The active area of the interferometers as determined by A-B oscillations is substantially smaller than the lithographic or top gate area, which is consistent with the large distance of the 2D gas from the surface and the large negative voltages applied to the surface gates. With the 2D gas 200nm from the surface, at nominal full depletion of the electrons below the gate the border of the active area must be 200nm internal from the lithographically defined gate structure. Since the gates are typically operated at much larger negative voltages than full depletion (~ -2V), the active area will be reduced substantially further. Even with this large bias and consequent wide lateral depletion profile, the d.c. transport through the device shows survival of 5/2, 7/3, and 8/3 fractional states: this may be due to the relatively high electron densities used in this study. The density used here is substantially higher than in other studies [6-7]. In our own experiments using samples at their lower densities and without illumination, only a resistive lump is observed between filling factors 2 and 3, with no resolution of fractional states. Since both higher density and sample illumination are used in our studies, either or both may be the origin of the higher quality transport observed in our samples.

## SUPPLEMENT REFERENCES:

[1] Willett R-L, Pfeiffer L-N, West K-W (2009) Measurement of filling factor 5/2 quasiparticle interference with observation of e/4 and e/2 period oscillations, *Proc Natl Acad Sci* **106**, 8853.

[2] Willett R-L, Pfeiffer L-N, West K-W (2010) Alternation and interchange of e/4 and e/2 period interference oscillations consistent with filling factor 5/2 non-Abelian quasiparticles, *Phys Rev B* **82**, 205301.

[3] Willett R-L, Manfra M-J, Pfeiffer L-N, West K-W (2007) Confinement of fractional quantum Hall states in narrow conducting channels, *Appl Phys Lett* **91**, 052105-052107.

[4] Rosenow B, Halperin B-I (2007) Influence of Interactions on Flux and Back-Gate Period of Quantum Hall Interferometers. Phys. Rev. Lett. **98**, 106801-106804.

[5] Halperin B-I, Stern A, Neder I, Rosenow B (2011) Theory of the Fabry-Perot quantum Hall interferometer, Phys. Rev. **B**83, 155440.

[6] Zhang Y, McClure D-T, Levenson-Falk E-M, Marcus C-M, (2009) Distinct signatures for Coulomb blockade and Aharonov-Bohm interference in electronic Fabry-Perot interferometers, Phys. Rev. **B**79 241304(R).

[7] Godfrey M-D, Jiang P, Kang W, Simon S-H, Baldwin K-W, Pfeiffer L-N, West K-W, (2007) Aharonov-Bohm-Like Oscillations in Quantum Hall Corrals. arxiv.org/abs/cond-mat/0708.2448.

**FIGURE CAPTIONS:**

Figure S1: A) magneto-transport through interferometric device of nominal area 1, demonstrating $R_L$ from near filling 4 to 2. The prominent fractional quantum Hall states at 7/2 and 5/2 are marked. B) small amplitude magneto-resistance oscillations (period $\Delta B_2(7/2)$) near 7/2 filling factor. The period is roughly 73 Gauss. The corresponding A-B oscillations that determine area are shown in the lowest panel of main text Figure 3. C) product of $\Delta B_2(7/2)$ and device area versus device area, comparing to the data from 5/2 filling. The expected value of this product for 7/2 is shown as the lower solid line in the figure, and the measured value (blue symbol) is in coarse agreement.

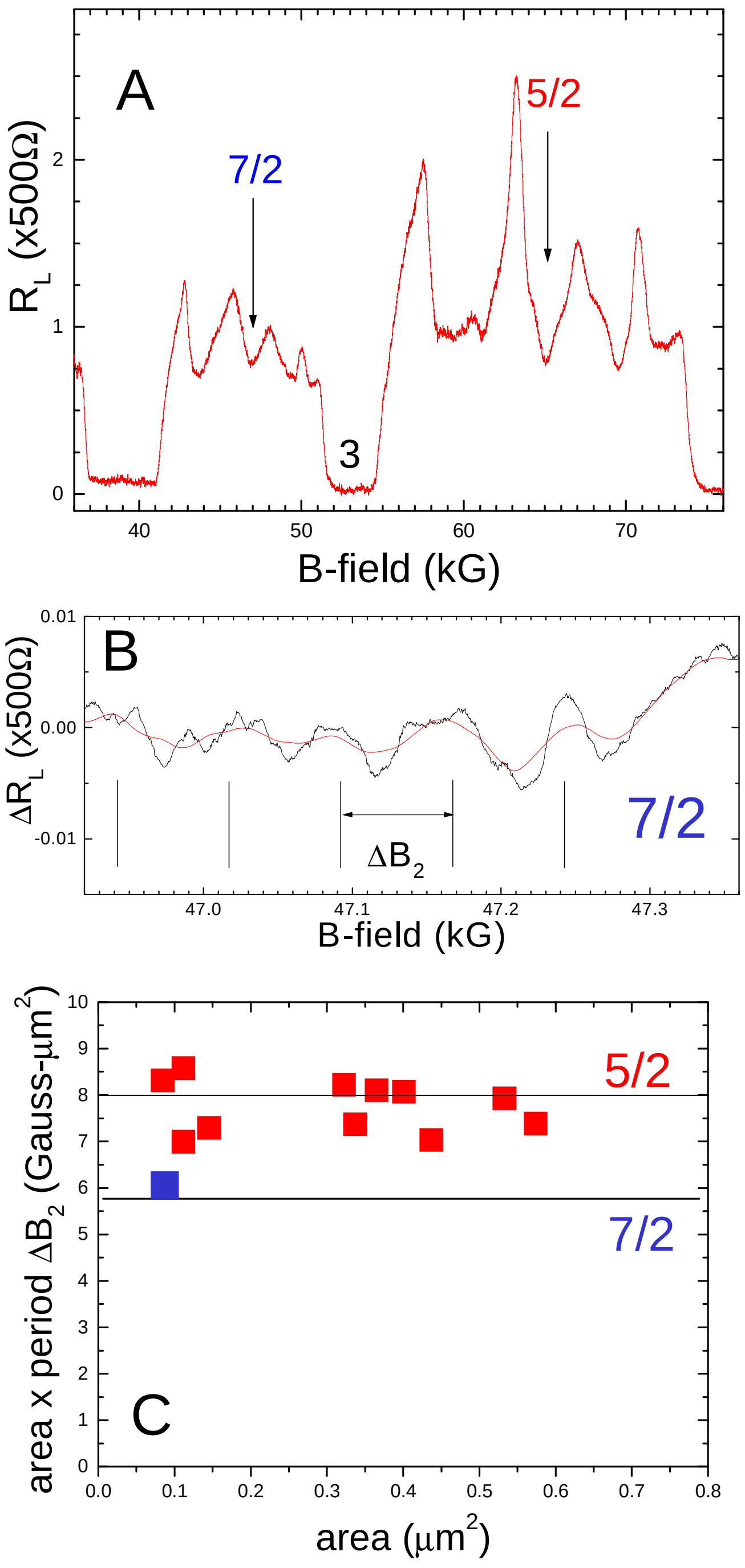

Figure S1